\documentclass[fleqn,11pt]{article}

\usepackage[dvips]{graphicx} \usepackage{color}

  \def\NN{\hbox{I\kern-.2em\hbox{N}}}
\def\RR{\hbox{I\kern-.2em\hbox{R}}}

\title{Quantum Monte Carlo activation barrier for hydrogen dissociation on copper to unprecedented accuracy.}

\author{Philip E. Hoggan}

\begin{document}

\maketitle

\thispagestyle{empty}

\centerline{Institut Pascal, UMR 6602 CNRS\@.}

 \centerline{ LabEx IMobS3}

\centerline{Campus Universitaire des C\'ezeaux}

\centerline{4 Avenue Blaise Pascal, TSA 60026, CS 60026}

\centerline{ 63178 AUBIERE Cedex, France.}

\vskip1mm


\begin{abstract}

 Many chemical reactions involve bond-dissociation. This is also true for reactions at solid surfaces, in which the dissociation step is often limiting but facilitated in comparison to gas phase reaction channels.
This work considers H$_2$ dissociation. Reliable molecular beam results are available for this reaction at some copper surfaces.

Heterogeneous catalysis by copper is simulated. It was investigated in our previous work since it is in many ways a prototype metal presenting a close-packed surface here.

These hydrogen molecules are adsorbed at Cu(111) and fixed geometries on the dissociation reaction pathway for stretched and distant equilibrium H$_2$ are given by using Density Functional Theory (DFT) calculations in a plane wave basis.

The PBE wave-functions at these bond-lengths serve as trial input for Quantum Monte Carlo (QMC) simulations of the ground states to obtain highly accurate correlated results for the associated activation barriers indicating the catalytic effect on this dissociation. This correlation varies as bonds dissociate, requiring its accurate evaluation.

Finite size effects and fixed-node error are possible limitations to accuracy of this type of QMC study.

We are able to limit fixed node error, using certain trial wave-functions. The finite size effect is considerable, although comparing two adsorbed  geometries cancels about 90\% with respect to clean surfaces.

The pseudo-potential used to represent the atomic core of copper must also be determined carefully: we leave 11 active electrons including the 3d shell for an Ar-core pseudo-potential.

\end{abstract}
\vskip5mm

\newpage

\section{Introduction}

The driving force behind using the Quantum Monte Carlo (QMC) simulation approach, rather than Density Functional Theory (DFT) comes from the need to describe phenomena that are not adequately catered for by using DFT which is three of orders of magnitude quicker.

The present study simulates heterogeneous catalysis that enhances bond dissociation. This step is frequently the initial (and often limiting) step of an industrial reaction (see, for example \cite{Diaz}). Bond dissociation is difficult to describe using most quantum theory approaches for isolated diatomic molecules. Taking electron correlation into account does lead to the prediction of the observed products. Therefore, heterogeneous catalysis requires an approach that can correctly include almost all electron correlation (is varies as bonds dissociate), in addition to the interactions involved.
\vskip2mm
 It is well-known that Hartree-Fock methods fail to describe bond dissociation correctly. This is so, even for homo-nuclear diatomics. Better results need lengthy extensive Configuration Interaction (CI) to cater for electron correlation. DFT methods present a rapid computational alternative, which performs better in the dissociation limit when certain functionals are used. In the interests of comparing {\it ab initio} approaches, the Perdew, Burke, Ernzerhof (PBE) functional was used in this work. It gives a reasonable gas phase barrier, however, on Cu(111) limiting a relaxed compact cubic structure, the barrier is found to increase significantly, as opposed to observed catalytic effects. \cite{Nruth}
 \vskip4mm
In essence, there is no alternative than to include correlation and to begin with a trial wave-function that behaves as correctly as possible, in particular close to dissociation. The transition state for H$_2$ dissociation on Cu(111) is our test case and it has a very stretched bond in the input geometry, since the transition state is quite 'late'. This means that it occurs when the bond distance is already much greater than the equilibrium value (i.e. 0.7 {\AA}). Transition state geometry poses a problem in QMC simulations since geometry optimisation is prohibitively long because QMC force-constant evaluation is very time-consuming. Since experimental geometries are scarce in this case, it is usual to adopt DFT geometries, if possible validated by empirical fitting to reliable measured barrier heights \cite{Diaz}). It is applied to obtain indications of relative barrier heights which correspond well to heterogeneous catalysis (see \cite{Nruth}). Equilibrium H$_2$ at 6.5 \AA \hskip1mm from Cu(111) gives the QMC reference.

\vskip4mm
 This work shows that Quantum Monte Carlo simulations at a Cu(111) surface of a relaxed bulk Face-Centered Cubic (FCC) lattice gives the observed behaviour. Electron correlation is uniquely well accounted-for. The corresponding energy contribution varies dramatically during adsorption and reaction, therefore it must be determined exactly.

 A set of DFT benchmarks by Thakkar \cite{Thakk}  highlights the poor performance of 11 much used DFT functionals for evaluating electron correlation energy.

 A comparative study of Diffusion Monte Carlo (DMC) and DFT-MRCI (DFT-Multi-reference configuration interaction) methods for excitation energies tests similar cases to the present rate-limiting activation barriers (in system size and electron re-arrangement): percentage errors in the total excitation energies are 3 for DFT-MRCI and only 0.4 for DMC. (See Lester {\it et al} \cite{Les}).

\vskip2mm
The Quantum Monte Carlo approach uses statistical physics over a large population of configurations i.e. a set of instantaneous particle positions in space. Often configurations are called 'walkers'. After equilibration, for numerous data-points N, high accuracy is obtained with the error decreasing as $1 \over \sqrt{N}$.
\vskip2mm
  Trial wave-function quality is carefully optimised.and finite size effects catered for.  The CASINO code is used, which is well-suited to periodic solids and can be made to scale slowly with system size (as n$^3$ for n electrons), by expanding the plane-wave basis in cubic splines (blips) \cite{casino}.

 A pre-requisite for QMC simulations is access to a super-calculator e;g. Bluegene (BG/Q) see \cite{prace}.
\vskip2mm

 To limit finite-size effects, a large k-point grid should be used. This was limited by computer memory to one much smaller than the converged DFT/plane wave grid.
 In this work, we were able to reach a 3 3 1 grid using the 28GB RAM/node of MareNostrum III. This grid was used to test corrections to finite-size effect. The DFT calculations converge only at 16 16 1 or higher. Twist-averaging is used to correct for finite size error. A high proportion (around 9/10) of this error cancels between the Transition State (TS) and asymptotic geometry. Production runs used a 2 2 1 grid which further reduces finite size error by a factor 25 (127 for 3 3 1) c.f a 1-point calculation.
\vskip2mm

 The solid is a 2 by 2 four layer copper slab (16 atoms). Each atom has 11 active electrons including 3d electrons (that are dense in the core-region) since the Troullier-Martins pseudo-potential used has an Ar-core. DFT/plane wave calculations with the pseudo-potentials were carried out with ABINIT. This software is available online \cite{abinit}.

 A TS geometry from a specific DFT fit to experimental data and an asymptote with equilibrium H$_2$ at long distance (6.5 \AA) are compared by QMC. The energy difference gives the activation barrier.

 Neither the stretched H$_2$ nor asymptote trial wave-function is of good quality. This is manifest since the first steps of the DMC simulation using them shows a surge in population followed by an excessively time-consuming equilibration. These phenomena should be attenuated.
\vskip2mm

 The situation must be improved by carefully optimising the Jastrow factor \cite{jastrow} (essential work, taking up to 5\% of the total time). This factor uses a polynomial expansion in the variables of explicit correlation. Its product with a Slater determinant gives the trial wave-function.

 It accounts for electron-electron terms, electron-nucleus terms and also the very numerous electron pair-nucleus terms (e-e, e-N and e-e-N). In previous work, the merits of pre-correlated trial wave-functions using simple explicitly correlated geminals or the related plane-waves has been proposed \cite{PEH,Alavi}. In this work, we use the linear Jastrow optimisation directly on trial wave-functions. This optimisation is carried out during a relatively rapid Variational Monte Carlo (VMC) stage of the calculation. A suitable choice of optimisation method is essential. There are two efficient approaches, the first is a linear method, currently applicable in CASINO to real wave-functions in a plane-wave basis. The other is a new approach, \cite{casino} but which generates a huge parameter set for this copper slab model. A good compromise is the linear technique with a reasonably high expansion order of 5 and unrestricted spin states. Surface Cu is optimised separately.

\section{Methods}
 As mentioned above, DFT\@ is not suitable for stretched bonds nor adsorbtion (which involves long-range interactions).
A method is required that gives
almost all the electronic correlation. The ideal choice is the quantum
Monte Carlo (QMC) approach. In variational quantum Monte Carlo (VMC)
correlation is taken into account by using a trial many-electron wave
function that is an explicit function of inter-particle distances.
Free parameters in the trial wave function are optimised by minimising
the energy expectation value variationally.  The trial wave functions used here are of Slater--Jastrow form, consisting of Slater determinants
(orbitals taken from Hartree--Fock or DFT codes), multiplied by geminals and a
Jastrow factor including electron pair and three-body
(two-electron and nucleus) terms.

A second, more accurate step, takes the optimised Jastrow factor as
data and carries out a diffusion quantum Monte Carlo (DMC) calculation, based on
transforming the time-dependent Schr\"{o}dinger equation to a diffusion
problem in imaginary time. An ensemble in configuration space is propagated in this imaginary time-variable to obtain a highly accurate ground state.

\section{Setting up the model system.}

For applications to heterogeneous catalysis, a slab of the metal is constructed. It must be much longer than the maximum bond-length and sufficiently thick for the surface layer perturbation during geometry optimisation to be attenuated at a depth within the slab to the extent that the bulk parameters apply to the final layers. The geometry must be fixed for the input to QMC and the best practice for this is to take experimental values if they are available, or failing that, a PBE value obtained by stretching the bond in H$_2$. The H$_2$ is 2 {\AA} above the surface and horizontal.

 A super-cell is thereby defined by repeating the slab in the direction perpendicular to the surface. This cell is quite large (over 700 electrons, after the use of suitable pseudo-potentials) but QMC is rendered linear with system size by expanding the plane-waves in cubic splines or B-splines (blips).

\section{Trial wave-function and choice of pseudo-potential}

The trial wave-function should be compact and have low variance. Consider a plane-wave basis and the transition state or asymptotic geometry surface-adsorbate systems studied in the present study. It is readily observed that Troullier Martins pseudo-potentials which are relatively hard result in some compactness of the blip wave-function.

  However, the present copper slab wave-function has a high variance of 0.2 Ha/electron, halved by careful optimisation of the Jastrow factor (see below). One cause of this poor initial condition is that pseudo-potentials are difficult to construct when atoms possess d-electrons that penetrate the region of space occupied by the core whilst also being diffuse and therefore of significant electron density in the outer region of the atom where the valence electron density is high. The 3d electrons are explicitly valence. Variance is much lower for Pt slabs. A Pt core is quite free of d-density.\cite{PEH}

 A short time-step is necessary for the Diffusion Monte Carlo stage of the simulation, that provides the almost exact ground state of the combined system, in a fixed-node short time-step approximation.
Furthermore, since is copper is a heavy atom, this time-step can be expected to be particularly short and the value is given an initial value of 0.009 au.

\section{Finite size effects.}

 Adsorbed systems generally involve reference to a clean metal surface. This implies that extensively delocalised states are involved, which formally describe conducting bands of electrons. In DFT treatments of metals using software based on periodicity, i.e. with a basis of plane waves, work in the first Brillouin zone shows that these states and the surface energy converge slowly with the size of k-point grid.

The present state of the art QMC for solid state applications use a modest size of grid because it must be 'unfolded' for QMC. This means the number of particles is repeated the same number of times as there are k-points. Often, therefore, the QMC calculation is limited in practice to a grid that is small with respect to convergence, (In practise, the wave-function input data must fit the available computer memory. This work used a 2 2 1 grid, where the atoms in the cell and electrons are already repeated four times, giving 712 electrons). This does improve the energy error in surface formation for clean metals, as shown by previous work on copper \cite{PEH} by over a factor 25, compared with the usual single k-point but a short-fall with respect to convergence remains.

This phenomenon is referred to as the finite size effect. It may be greatly reduced by twist-averaging and some further analytical corrections for the asymptotes \cite{FES}. At each twist of the 16, re-equilibration is followed by collecting 100 000 data-points.

In this work, the possibility of eliminating the clean metal surface from the reference for activation barrier calculation is investigated. The idea of referring to adsorbed reactants to limit the influence of fully delocalised conduction bands is tested. As seen in \cite{PEH} , surface states which are dominated by a 2D symmetry are less prone to finite size effects, this strategy represents a definite progress, however, overall error is less than 2mHa and due to the essentially 3-D delocalised states involved.

\section{Quantum Monte Carlo Simulation methods and application.}

\subsection{Variation Monte Carlo}

A preliminary Variational Monte Carlo (VMC) calculation is carried out in order to generate several thousand
configurations (instantaneous points in the direct space of the electrons). VMC is driven by energy minimisation. The
'local energy' $H \psi / \psi$ is evaluated, including kinetic energy terms that are smoother and have lower variance
in exponentially decaying bases, as shown in work on wave-function quality \cite{stoqmc}.
Electron correlation is introduced via a Jastrow factor which can be optimised by Variational Monte Carlo methods.

This VMC optimisation used 20480 configurations on 2048 cores, the linear Jastrow optimisation method was chosen after an initial variance minimisation step.

\vskip2mm This optimisation procedure rapidly generates a data file containing the optimised numerical parameters for
the electron-electron and electron pair-nuclear contributions to the Jastrow factor.

A final VMC calculation generates the initial configurations required for the Diffusion Monte Carlo step(DMC) also
10-20 per core, typically. The previous VMC steps need to generate at least as many configurations.

\subsection{Diffusion Monte Carlo}

 In
the DMC method the ground-state component of the trial wave function is projected out by solving the Schr\"{o}dinger
equation (SWE) in imaginary time.  This is accomplished by noting that the imaginary-time SWE is a diffusion equation
in the $3N$-dimensional space of electron coordinates, with the potential energy acting as a source/sink term.  The
imaginary-time SWE can therefore be solved by a combination of diffusion and branching/dying processes.  The
introduction of importance sampling using the trial wave function transforms the problem into one involving drift as
well as diffusion, but greatly reduces the population fluctuations due to the branching/dying process.

The Fermionic
antisymmetry of the wave function has to be maintained by constraining the nodal surface to equal that of the trial
wave function.

\subsection{Application of QMC: Numerical methods and algorithms used.}

This suitable reduction to relatively few carefully selected active electrons  allows an accurate wave-function representation of the
complex system involved, thus lending itself to the use of QMC to account almost exactly for electron correlation.

The integration is performed using a Monte Carlo method, with the Metropolis algorithm being used to sample
electronic configurations distributed as the square of the trial wave function.

 The DMC method was then used to determine a highly accurate value for the ground-state energy.  The computational
effort required by the DMC calculations is very much greater than that required by the preliminary DFT or VMC
calculations, dominating the total computer time required.

BG/Q runs on 8192 cores with 10 configurations per core as the target weight have become routine. OpenMP also allows hybrid parallel strategies to be used with 4 threads running per core.

This approach led to calculating DMC energies for various DFT-generated atomic configurations, the activation energy barrier height was
obtained for a model hydrogen bond dissociation reaction taking place at the Cu(111) surface. The physics underpinning the chemical process of
catalysis was described.


The value of 1-2 mili-Hartree is retained as
threshold for significant energy differences to be resolved by QMC. It appears to be similar to that of systematic errors in this QMC work that we strive to eliminate in future.

\subsection{The need for QMC and its originality}

Reference \cite{foulkes} provides an overview of the VMC and DMC methods.

These carefully selected electrons allow an
accurate wave-function representation of the complex system involved,
thus lending itself to the use of QMC to account accurately for
electron correlation. To our knowledge, our previous work \cite{PEH} is the first time that
QMC was used to study a reaction taking place on a surface between co-adsorbed molecules. This novel approach is now proposed with improvements made in the present work.
\vskip4mm
The initial interaction energy from DFT work is 0.1 a.u., i.e 2.7 eV,
but differences as low as 0.1--0.2 eV may be significant,
corresponding to weak (dispersion or polarisation)
interactions. This is the value retained as threshold for significant energy differences to be resolved by QMC.

The statistical error bar $\Delta$ on the QMC total energy must be
small compared with the energy difference to be resolved.  Assuming
the cost of the equilibration phase of a QMC calculation is
negligible, the statistical error bar falls off as $1/\sqrt{T}$, where
$T$ is the computational effort in core-hours.

To sum up, the accuracy for a given wall-time depends on the quality of the trial wave-function and overall accuracy essentially increases with the square-root of the number of data-points collected.

\section{Results.}

The QMC results for the hydrogen dissociation activation energy barrier on Cu{111) are quoted with statistical error, to which a systematic error estimated at 1.5 kcal/mol should be added (based on the variance of the trial wave-function). The raw number is the internal energy prior to correction for finite size-effects not included in the twist-averaging process. The activation barrier, corrected in this way, is called c-QMC. The experimental value is an empirical DFT value fitted to the best molecular beam measurements.
\vskip10mm
\subsection{Estimated barrier heights in kcal/mol}
\begin{tabular} {|c|c|c|c|c|}
 \hline
Raw QMC & 15.00  (se 0.55) & c-QMC & 14.79  (se 0.55)  & Exp$^t$ 14.48 \\

 \hline
\end{tabular}
\vskip5mm
The final figure is 14.8 +/- 1.64 compared to 14.5 (molecular beam value) in kcal/mol.

This measured value could be taken as a lower bound for the activation barrier within the activated complex model. The actual system will depend on temperature, surface re-arrangement and possible defects. Nevertheless, agreement is perfect and the systematic error on the QMC value may be reduced by methods currently being investigated.

\section{Note}

These results are from data files dated at the end of may 2014. This report is a snapshot of the QMC simulation described here at that time.

\section{Acknowledgements}

The data for TS structure is from reference \cite{Diaz}.

Computer time was obtained within PRACE call 6 on the MareNostrum machine, at BSC, Barcelona. Runs began in october 2012 and consumed about 6 million BG/Q equivalent core-hours. We are grateful to PRACE for enabling this project to be backed. We acknowledge help from CASINO authors, N. D. Drummond and M. D. Towler during installation and tests on the platform.

\end{document}